\documentclass[smallabstract,smallcaptions]{dccpaper}
\usepackage{blindtext}
\usepackage{amssymb}
\usepackage{amsmath}
\usepackage{fancyhdr}
\usepackage{graphicx}
\usepackage{float}
\usepackage{algorithm}
\usepackage{algpseudocode}
\usepackage{subcaption}
\usepackage{multirow}
\usepackage[algo2e]{algorithm2e}
\usepackage{bm}

\long\def\comment#1{}

\newfont{\bbb}{msbm10 scaled 700}

\newfont{\bb}{msbm10 scaled 1100}

\renewcommand{\arg}{{\hbox{arg}}}

\usepackage{color}
\usepackage{url}

\begin{document}
\title{\large
\textbf{Fractional Motion Estimation for Point Cloud Compression}}

\author{Haoran Hong$^{\star}$, Eduardo Pavez$^{\star}$, Antonio Ortega$^{\star}$,
Ryosuke Watanabe$^{\dagger}$, \\[0.5em] Keisuke Nonaka$^{\dagger}$ \\[0.5em]
\thanks{This work was funded in part by KDDI Research, Inc.~and by the National Science Foundation (NSF CNS-1956190).  }
$^{\star}$University of Southern California, Los Angeles CA. 90089 USA \\
$^{\dagger}$KDDI Research, Inc., Japan\\
$^{\star}$\url{{haoranho,pavezcar,aortega}@usc.edu} \\ $^{\dagger}$\url{{ru-watanabe,ki-nonaka}@kddi-research.jp}
}

\maketitle
\begin{abstract}
Motivated by the success of fractional pixel motion in video coding, we explore the design of motion estimation with fractional-voxel resolution for compression of color attributes of dynamic 3D point clouds. Our proposed block-based fractional-voxel motion estimation scheme takes into account the fundamental differences  between point clouds and videos, i.e., the irregularity of the distribution of voxels within a frame and across frames. We show that motion compensation can benefit from the higher resolution reference and more accurate displacements provided by fractional precision. 
Our proposed scheme significantly outperforms comparable methods that only use integer motion. The proposed scheme can be combined with and add sizeable gains to state-of-the-art systems that use transforms such as Region Adaptive Graph Fourier Transform and Region Adaptive Haar Transform.

\end{abstract}

\section{Introduction}
\label{sec:introduction}
Recent progress in 3D acquisition and reconstruction technology makes the capture of 3D scenes ubiquitous. 
%
In dynamic point clouds, each frame consists of a list of data points with 3D coordinates and RGB color values. Since point clouds in raw format would require a huge amount of bandwidth for transmission there has been a significant interest in point cloud compression techniques, which has led to MPEG standardization efforts  \cite{overview2019}, considering both  video-based point cloud compression (V-PCC) and geometry-based point cloud compression (G-PCC)  \cite{overview2019,overview2020}.



Methods for inter-frame (temporal) prediction have been proposed to achieve efficient compression of dynamic point clouds. These methods can be grouped into three main categories. 
In \textit{voxel-based} schemes \cite{Dorina2016}, where a motion vector (MV) is estimated for each voxel, a few points in both the prediction and reference frames are selected as anchors to establish correspondence via spectral matching, leading to a set of sparse MVs. Then, using a smoothness constraint, a dense set of MVs can be obtained from the sparse set to provide motion for all remaining points. 
In \textit{patch-based} techniques \cite{Xu2020}, motion estimation (ME) is considered as an unsupervised 3D point registration process wherein a MV is estimated by iterative closest point (ICP) \cite{ICP} for each patch generated by K-means clustering. 
In this paper we focus on \textit{block-based} methods, where frames to be predicted are partitioned into several non-overlapping 3-dimensional blocks of a given size. For each block, the best matching block in a reference frame is selected according to specific matching criteria, which can be based purely on geometry, e.g., an ICP-based approach that generates rigid transforms \cite{Rufael2017}, or can use a combination of geometry and color attribute information \cite{Ricardo2017}.
Recent work has also focused on block-based motion search speedup, including both efficient search pattern design and search window reduction \cite{Camilo2018,Camilo2019,Souto2020,SantosICIP2021}.


Our work is motivated by the observation that ME with sub-pixel accuracy is an essential tool for modern video coding \cite{Girod1993}, while   
all the aforementioned ME methods for dynamic point clouds are based on integer-voxel displacements. 
There are two main reasons why a direct extension of video-based fractional ME to 3D contexts is not straightforward.
First, point clouds are irregularly distributed within each frame, i.e., only those voxels that correspond to the surfaces of objects in the scene contain attribute information. Thus, while interpolation of attributes at new voxel locations can be based in conventional methods, we have the additional challenge of choosing only those new voxel locations that are consistent with object surfaces, even though those surfaces are not explicitly known.   
For example, we would like to avoid creating additional, fractional accuracy voxels 
\textit{inside} an object.      
Second, voxels are inconsistent across frames, i.e., both the number of voxels and their distribution in space are different from frame to frame. 
Thus, since two matching blocks in consecutive frames will in general have a different number of voxels containing attribute information, we will need to develop alternatives to the one-to-one pixel (or sub-pixel) matching commonly used for conventional video. 

In this paper, we focus on fractional-voxel motion estimation (FvME) under the assumption that  integer-voxel MVs (IvMVs) have already been obtained using an existing integer-voxel motion estimation (IvME) scheme 
\cite{Ricardo2017,Camilo2018,Camilo2019,Souto2020}. 
Specifically, in this paper we use precomputed IvMVs from a public database \cite{IMVdataset}.  
%
In our approach, we start by creating fractional voxels between pairs of \textit{neighboring} occupied integer voxels. Neighboring voxels are used to favor consistency with object surfaces, without requiring explicit estimation of the surfaces. Then, a higher resolution point cloud is obtained by interpolating attributes at each fractional voxel from the values at nearby integer voxels. 
FvME is implemented by searching fractional-voxel MVs (FvMVs) around the positions given by IvMVs 
and selecting 
the fractional displacement leading to the lowest motion compensation prediction error.
Motion-compensated prediction is implemented by directly copying, as the 
attribute for a voxel in a block in the current frame, the attribute of the \textit{nearest} voxel in the matched blocked in the reference frame.
Our proposed FvME scheme leads to improved performance over transform-based approaches without inter or intra prediction and is also significantly better than temporal prediction methods based on the IvMVs from \cite{IMVdataset}. 


\section{Fractional-Voxel Motion Estimation and Compensation}
\label{sec:FME}
%
%
\subsection{Motivation}
%
Real-world scenes and objects are captured by multiple calibrated and synchronized RGB or RGB-D cameras clusters from various viewing angles \cite{GROOT,8idataset}. 
After stitching and voxelization, dynamic point clouds are generated on integer grids.
%
%
Note that the 3D voxel coordinates are obtained as integer approximations to the ``true'' positions of the object in 3D space, while the optimal displacement between frames is unlikely to be exactly integer. Thus, a fractional voxel displacement can  be better than an integer-one, so that 
higher resolution MVs have the potential to provide more accurate motion and hence more accurate MC prediction.

Furthermore, distortion due to lossy coding in previously reconstructed point cloud frames can lead to higher prediction errors, while camera noise, lighting change in the capture environment, object movements, etc., may also result in noisy color attributes and in imperfect matches during motion search \cite{Zhao2017}. 
Thus, as for conventional video, where it is well known that fractional motion compensation contributes to noise removal, 
the process of generating higher resolution point clouds and attributes at fractional voxel locations can contribute to denoising and lead to improvements in the quality of the reference frames. 

%
\subsection{Occupied fractional voxels}
\label{sec:OFV}
In this section, we define fractional voxels and describe our proposed method for interpolation.
%
Based on the same design philosophy used for images and videos, fractional voxels are created at selected intermediate locations between voxels on the integer resolution grid. 
We define a fractional voxel of $1/2$ resolution (1/2-voxel), as a voxel at the mid point between any two neighboring integer-voxels.


As noted in the introduction, not all integer-voxels are ``occupied'' in 3D space, and those that are occupied typically correspond to object surfaces. Thus, in our proposed method, new fractional voxels are created only at locations in 3D space that are (approximately) consistent with the surfaces implied the location of occupied integer voxels  and  attributes are interpolated only at these newly created fractional-voxels. We say that two integer voxels with coordinates $v_j$ and $v_k$ are neighbors if their distance is below a threshold $\rho$. Then, a fractional voxel
is created only between neighbors $v_j$ and $v_k$ (assumed to be close enough so that they are likely to belong to the same surface) and the corresponding interpolated color attribute is computed as:
\begin{equation}
\begin{split}
&  C(v_i) = \frac{1}{2} \left( C(v_j) + C(v_k) \right), \\
& \text{with } \; L(v_i) = \frac{1}{2}(L(v_j)+L(v_k)) \;\; \text{and} \;\; \text{dist}(v_j,v_k) \leq \rho, v_j,v_k\in V_i,
\label{equ:identification}
\end{split}
\end{equation}
where $v_i$ is a voxel in the fractional-voxel set $V_f$ with color signal $C(v_i)$, $v_j$ and $v_k$ are voxels in the integer-voxel set $V_i$ with color signal $C(v_j)$ and $C(v_k)$, respectively. $L(\cdot)$ represents the coordinates of the voxel. $\rho$ is the distance threshold and  $\text{dist}(v_j,v_k)$ measures the Euclidean distance between the coordinates of $v_j$ and $v_k$.

\begin{figure}[htb]
\begin{subfigure}[t]{0.42\textwidth}
\includegraphics[width = 1\linewidth]{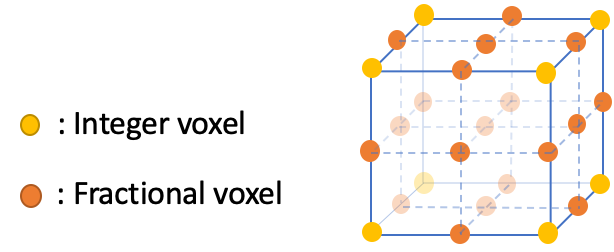}
\caption{}
\label{fig:halfvoxel}
\end{subfigure}
\begin{subfigure}[t]{0.53\textwidth}
 \includegraphics[width = 1\textwidth]{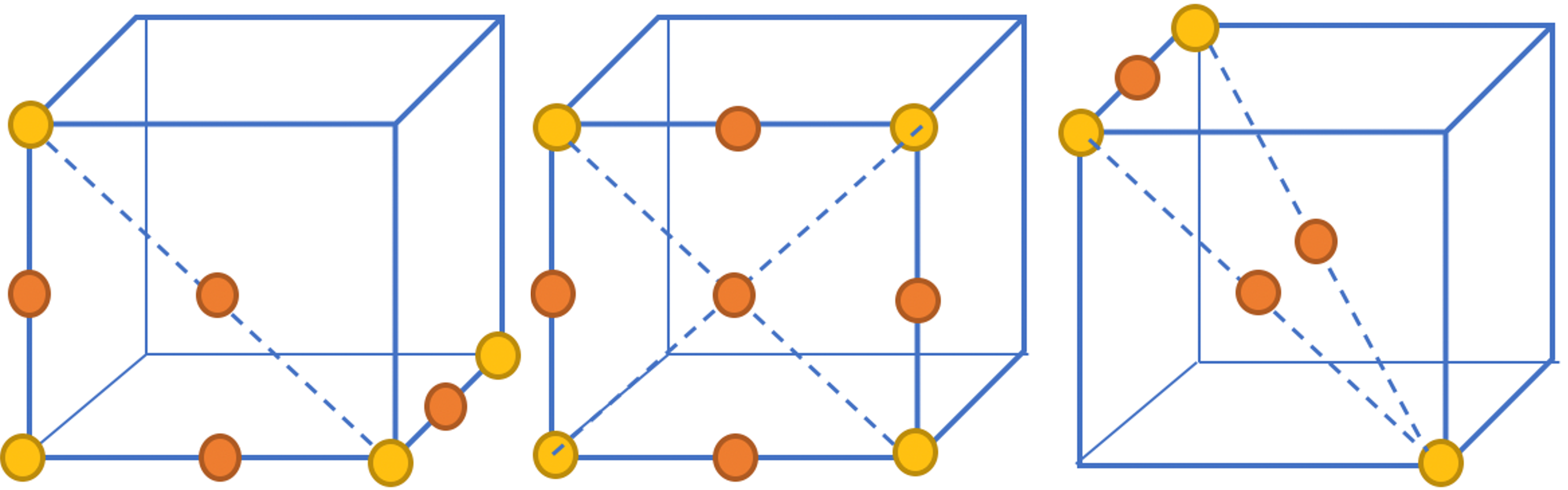}
\caption{}
\label{fig:identification}
\end{subfigure}
\caption{Integer and fractional voxels. Figure \ref{fig:halfvoxel} depicts all possible candidate integer and 1/2-voxels positions Figure \ref{fig:identification} shows 3 examples of occupied integer voxel positions with corresponding fractional voxels obtained from neighboring integer voxels. Note that these interpolated fractional voxels are more likely to belong to the same surface as the neighboring integer voxels they were obtained from.}
\end{figure}

Note that different pairs of integer voxels may produce the same fractional voxel. Thus, to remove repeated fractional voxels after interpolation,  attributes that belong to the same fractional voxel and are obtained by interpolation from different pairs of neighboring voxels are merged by averaging.
Fig.~\ref{fig:identification} shows several examples of possible fractional-voxels locations, where we can see that interpolation based on neighboring integer voxels tends to favor increasing the voxel resolution on the (implicit) surface where the voxels are located. 
%

\subsection{ME with fractional-voxel accuracy}

\begin{figure*}[h]
\centering
\includegraphics[width=1\textwidth]{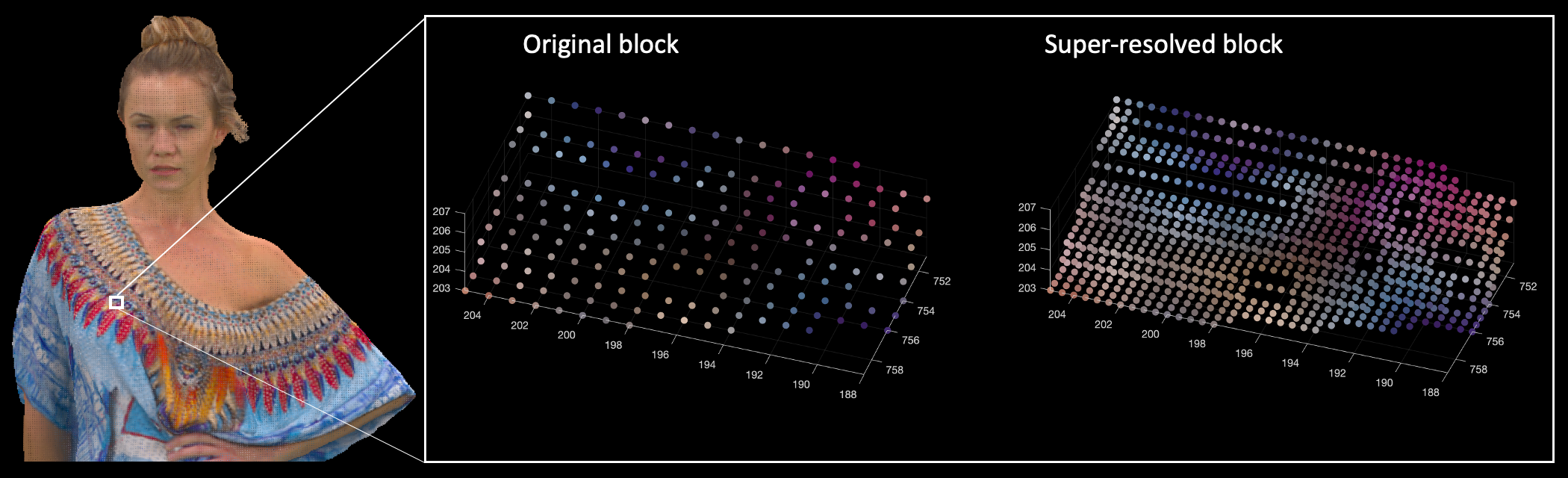}
\caption{Comparison between the original and super-resolved reference block.}
\label{fig:superresolution}
\end{figure*}

%
Due to the inconsistency of voxel distributions in consecutive frames, it is difficult to establish exact one-to-one correspondences between the voxels in two matching blocks. 
%
To generalize MC prediction for fractional motion in 3D space, 
we start by super-resolving the reference frame as described in Section~\ref{sec:OFV}. As we can see from Fig.~\ref{fig:superresolution}, the continuity among voxels and their corresponding attributes is significantly increased underlying surfaces, which provide better predictors when high resolution motion is available. The low pass filtering used for interpolation also contributes to attribute noise removal.

\begin{figure}[htb]
\begin{center}
\includegraphics[width = 1\linewidth]{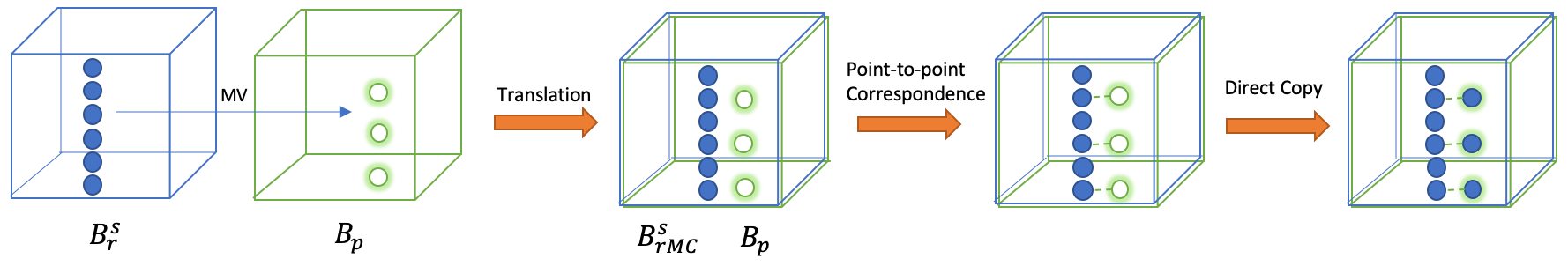}
\caption{Motion-compensated prediction.}
\label{fig:correspondence}
\end{center}
\end{figure}

Next, we estimate MVs in fractional precision for MC. The entire ME process is a coarse-to-fine procedure, including IvME and FvME. Each estimated MV is obtained as the sum of an IvMV and a FvMV displacement.
Assuming the IvMV $MV_i$ is given, the optimal FvMV $MV^{opt}_f$ is searched from a set of candidate fractional displacements.
Since we super-resolve the reference frame in 1/2-voxel precision, each coordinate of a fractional displacement $MV_f$ can take values in $\lbrace -\frac{1}{2}, 0, \frac{1}{2} \rbrace$, resulting in $27$ possible displacements. 
%
For a given fractional displacement $MV_f$, we predict each attribute in the current block from its nearest voxel in the translated super-resolved reference block, as depicted in Fig.~\ref{fig:correspondence}. 
Then the displacement with the smallest prediction error is chosen, that is,
\begin{equation}
\begin{split}
MV^{opt}_f & = \arg\min_{MV_f} \limits \sum_{v_i\in V(B_p)}{E_{pred}(C(v_i),C(v_{j'}))}, \\
 ~\text{s.t. } \; & j' = \arg\min_{j} \limits (\text{dist}(v_i,v_j)),v_j\in V(B_{rMC}^s), \\
 & L_b(B_{rMC}^s) = L_b(B_{r}^s) + MV,\\
 & MV = MV_f + MV_i,
\label{equ:FvME}
\end{split}
\end{equation}
where $B_r^s$ and $B_{rMC}^s$ represent the super-resolved reference block before and after translation with $MV$, respectively, $v_i$ and $v_j$ are voxels with color signals $C(v_i)$ and $C(v_j)$ in  blocks $B_p$ and $B_{rMC}^s$, respectively.  $E_{pred}(\cdot,\cdot)$ is the function for measuring the prediction error. $\text{L}_b(\cdot)$ represents the coordinates of the block while $\text{V}(\cdot)$ represents the set of voxels within the block.

\subsection{MC prediction with fractional-voxel accuracy}
Finally, we apply MC prediction using the obtained MVs in fractional precision. 
Specifically, once the voxels in the reference block are translated using the integer motion vector $MV_i$, they are further shifted by the obtained optimal fractional displacement $MV^{opt}_f$, as shown in \eqref{equ:FvME}.
Then, temporal correspondence are established from voxels in the predicted block $B_p$ to their nearest neighbours in the translated super-resolved reference block $B_{rMC}^s$ for motion-compensated prediction. The attribute of each voxel in the predicted block is predicted by copying the attribute of its corresponding voxel in the reference frame, that is,
\begin{equation}
\begin{split}
\forall v_i \in B_p \text{ , } C(v_i) = C(v_{j'}) 
& ~\text{s.t. } j' = \arg\min_{j} \limits (\text{dist}(v_i,v_j)),v_j\in B_{rMC}^s.
\label{equ:directcopy}
\end{split}
\end{equation}

\section{Experiments}
\label{sec:experiments}
\subsection{Dataset}
In this section, we evaluate the proposed FvME scheme for compression of  color attributes  of the  dataset of \cite{8idataset}, which consists of four sequences: \textit{longdress}, \textit{redandblack}, \textit{loot} and \textit{soldier}. Each sequence contains 300 frames.

Note that we assume  IvMVs are given and are used to  estimate FvMVs. Since IvMVs derived using different algorithms may lead to  different  FvMVs with disparate  coding performance, we start from  the publicly available 3D motion vector database  \cite{IMVdataset}. The IvMVs in \cite{IMVdataset} are selected to minimize a hybrid distance metric, $\delta = \delta_g + 0.35\delta_c$, which combines $\delta_g$, the average Euclidean distance between the voxels, and $\delta_c$, the average color distance in Y-channel\footnote{Note that the resulting IvMVs aim to select matching blocks with similar geometry ($\delta_g$) and color attributes ($\delta_c$) but there is no guarantee that this metric, and in particular the relative weight between the distances ($0.35$) is the optimal choice in terms of coding efficiency. Thus, it will be shown  in our  motion compensation experiments, that  these IvMVs can sometimes lead to performance below that of encoding methods that do not use motion compensation. In these cases performance can be improved by local refinement of the IvMVs from the database.}. 
We only consider motion for $16\times16\times16$ sized blocks. 
We implement a conventional inter-coding system where previously decoded frames are used as reference. 
\subsection{Experimental Settings}
Following  the MPEG Call for Proposal (CfP) for point cloud compression \cite{CTC2017}, we evaluate the proposed block-based FvME scheme (Proposed FvME) in groups of 32 frames, with the first frame coded in intra mode, and the rest coded using inter prediction. The threshold distance between integer voxels for interpolating fractional voxels is set to $\rho=\sqrt{3}$ in \eqref{equ:identification}.
Colors are transformed from RGB to YUV spaces, and each of the Y, U, and V channels are processed independently. When searching for the best candidate FvMV in \eqref{equ:FvME}, we use the squared distances to measure prediction errors. All blocks in the intra-coded frames undergo region adaptive graph Fourier transform (RA-GFT) \cite{RAGFT} while, in the inter-coded frames, all blocks are motion-compensated. After MC prediction, the residues are transformed using the graph Fourier transform (GFT) \cite{zhang2014point}. To compute the GFT, a threshold graph is built inside every block wherein voxels are connected to each other if their Euclidean distance is less than or equal to $\sqrt{3}$. If after thresholding, the resulting graph for a block is not connected, a complete graph is built instead, which results in the transformed coefficients consisting of a single approximation coefficient (DC) and multiple detail coefficients (ACs) for each block.
The DC coefficients of all blocks are concatenated and encoded together. Then, the AC coefficients  are coded block by block. This approach is equivalent to a single level RA-GFT \cite{RAGFT}. 
For all transforms, we perform uniform quantization and entropy code the coefficients using the adaptive run-length Golomb-Rice algorithm (RLGR) \cite{RLGR}. As for FvMVs overheads, since there are 27 FvMVs in total, 8 bits are used to signal each FvMV. For IvMVs, we use 4 bits to signal the value and 1 bit to signal sign for each axis and therefore, 15 total bits are used to represent an IvMV. The overheads of FvMVs and IvMVs are entropy coded by Lempel–Ziv–Markov chain algorithm \cite{LZMA}.

We considered the following schemes as baselines. IvME using the database motion (DM) for MC prediction and using DM with additional integer local refinement (DM+RF) for MC prediction. The local refinement uses different criteria that aims to minimize color errors only, instead of the hybrid errors used in the database. DM is refined by additional local search in integer precision to improve its matching accuracy over the original ones. The locally refined range for each axis is set to be $[-1,1]$, which entirely encloses fractional positions searched in the proposed FvME scheme. 

To evaluate the benefits of high resolution references and FvMVs, we propose two inter coding schemes which are using super-resolved reference blocks, with and without fractional motion vectors for compensated prediction. First, to evaluate the super-resolution method, we implement a scheme that considers IvME using integer local refined DM and super-resolved reference blocks for prediction, which is denoted by ``DM+RF+SR''. The difference between DM+RF and  DM+RF+SR is the resolution of the reference block.
Then, to evaluate benefits of FvMVs, we implement a  scheme that uses fractional resolution in both reference blocks and motion vectors, which is denoted by ``proposed FvME''.
For a fair comparison between inter coding schemes, all other test conditions are the same.
%

Additionally, to make our performance evaluation more complete, we include two state of the art (all intra) anchor solutions, namely, RA-GFT \cite{RAGFT} and region adaptive Haar transform (RAHT) \cite{raht}. For RA-GFT, block size 16 is used. The residues are entropy coded by RLGR.

\subsection{Evaluation Metrics}
The evaluation metrics are the number of bits per voxel (bpv) and average peak signal-to-noise ratio over Y component (PSNR-Y), 
%
\begin{equation}
PSNR_Y = -10 \log_{10} \left (\frac{1}{T} \sum_{t = 1}^T \frac{ \| Y_t - \hat{Y_t}  \|_2^2 }{ 255^2  N_t} \right), 
bpv = \frac{\sum_{t = 1}^T b_t}{\sum_{t = 1}^T N_t},
\label{equ:bpv}
\end{equation}
%
where $Y_t$ and $\hat{Y_t}$ represent original and reconstructed signals on the same voxels of $t$-th frame respectively, $T$ is the total number of frames, $b_t$ is the bits required to encode YUV components of $t$-th, including IvMVs and FvMVs overhead when necessary, and  $N_t$ is the total number of occupied voxels in $t$-th frame.
The Bjontegaard-Delta \cite{BDBR} results for bitrate (BD-rate) are also reported.

\subsection{Experimental Results and Analysis}

\begin{figure}[htb]
\begin{center}
\begin{subfigure}[t]{0.5\textwidth}
\includegraphics[width = \linewidth]{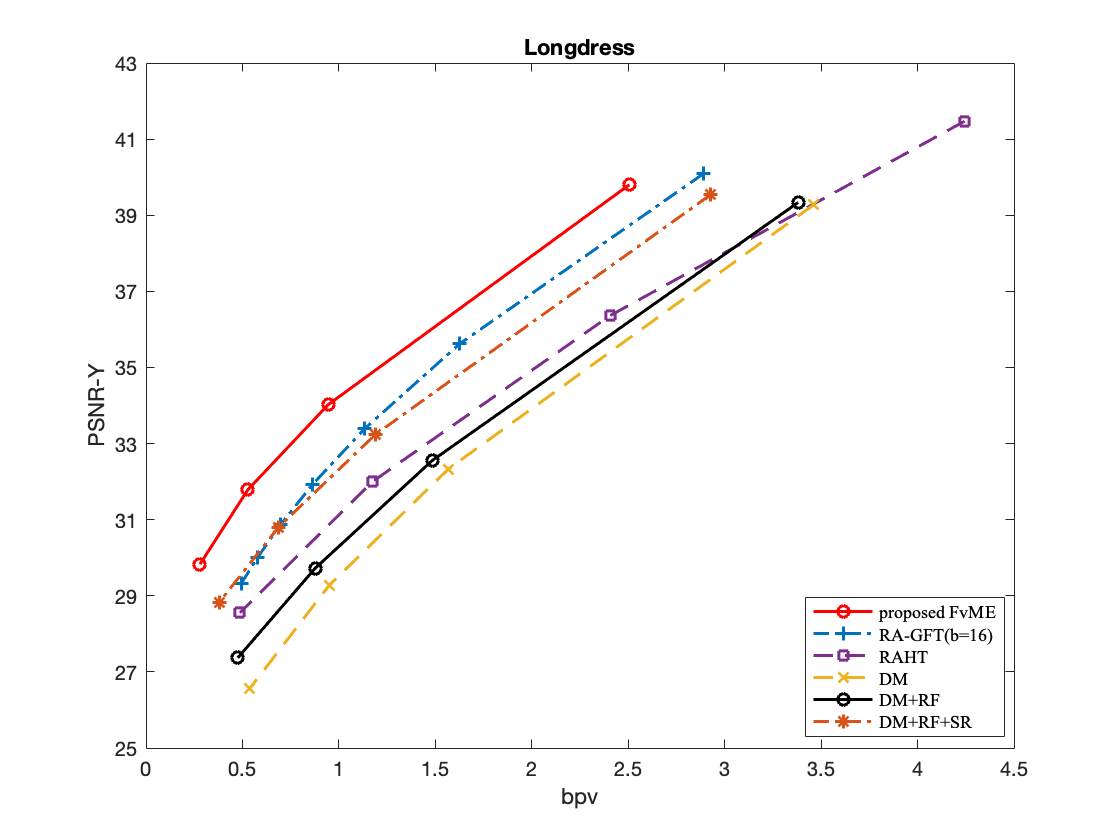}
\vspace{-6mm}
\caption{longdress}
\end{subfigure}
\hspace{-6mm}
\vspace{-1mm}
\begin{subfigure}[t]{0.5\textwidth}
\includegraphics[width = \linewidth]{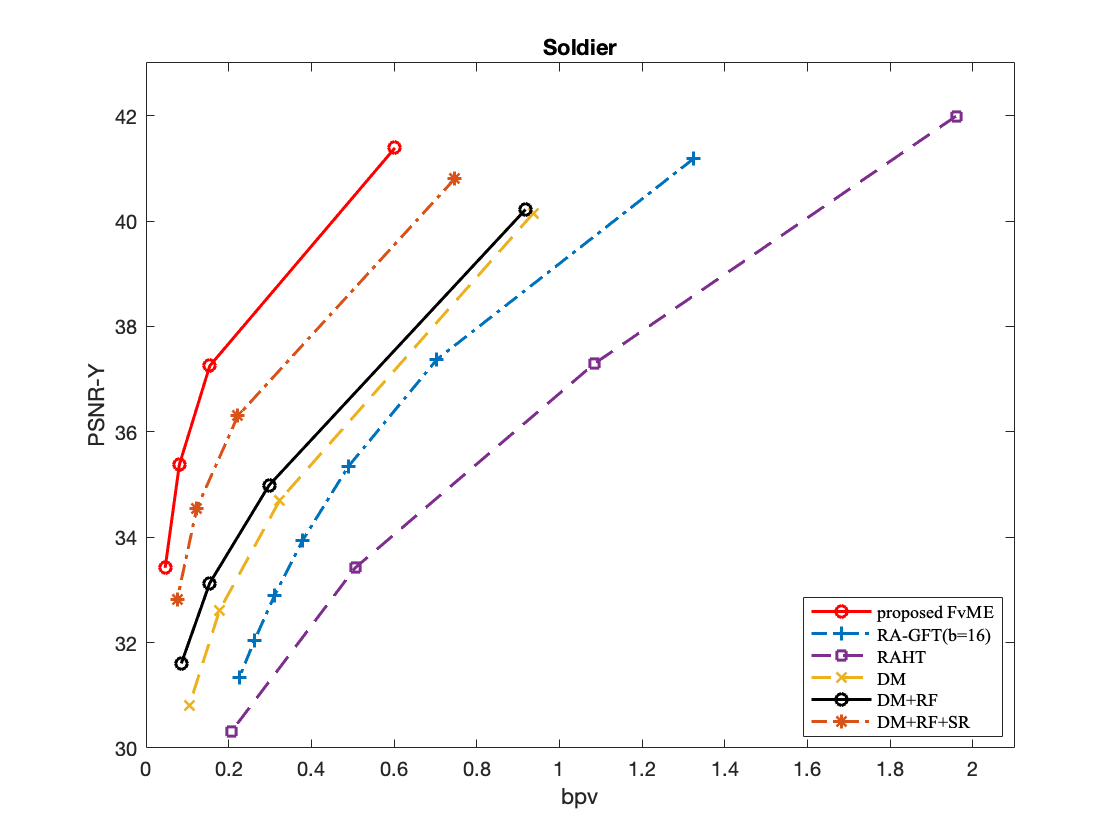}
\vspace{-6mm}
\caption{soldier}
\end{subfigure}
\begin{subfigure}[t]{0.5\textwidth}
\includegraphics[width = \linewidth]{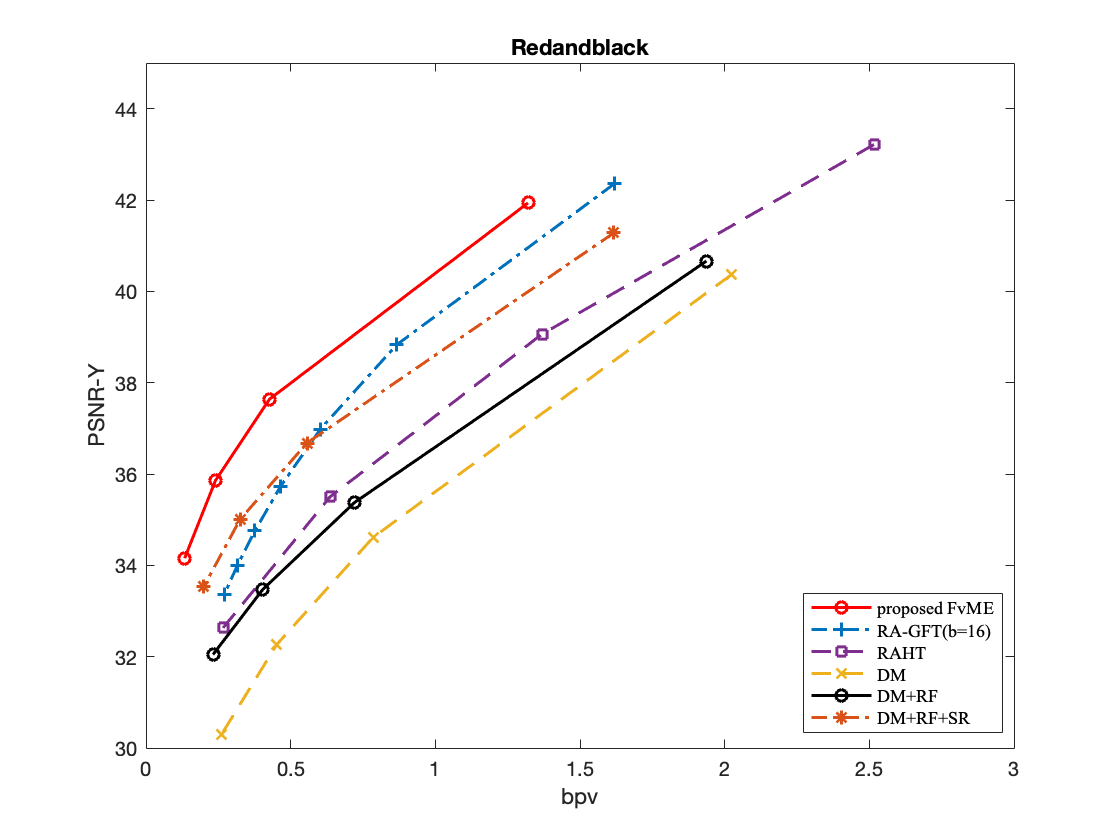}
\vspace{-6mm}
\caption{redandblack}
\end{subfigure}
\hspace{-6mm}
\vspace{-1mm}
\begin{subfigure}[t]{0.5\textwidth}
\includegraphics[width = \linewidth]{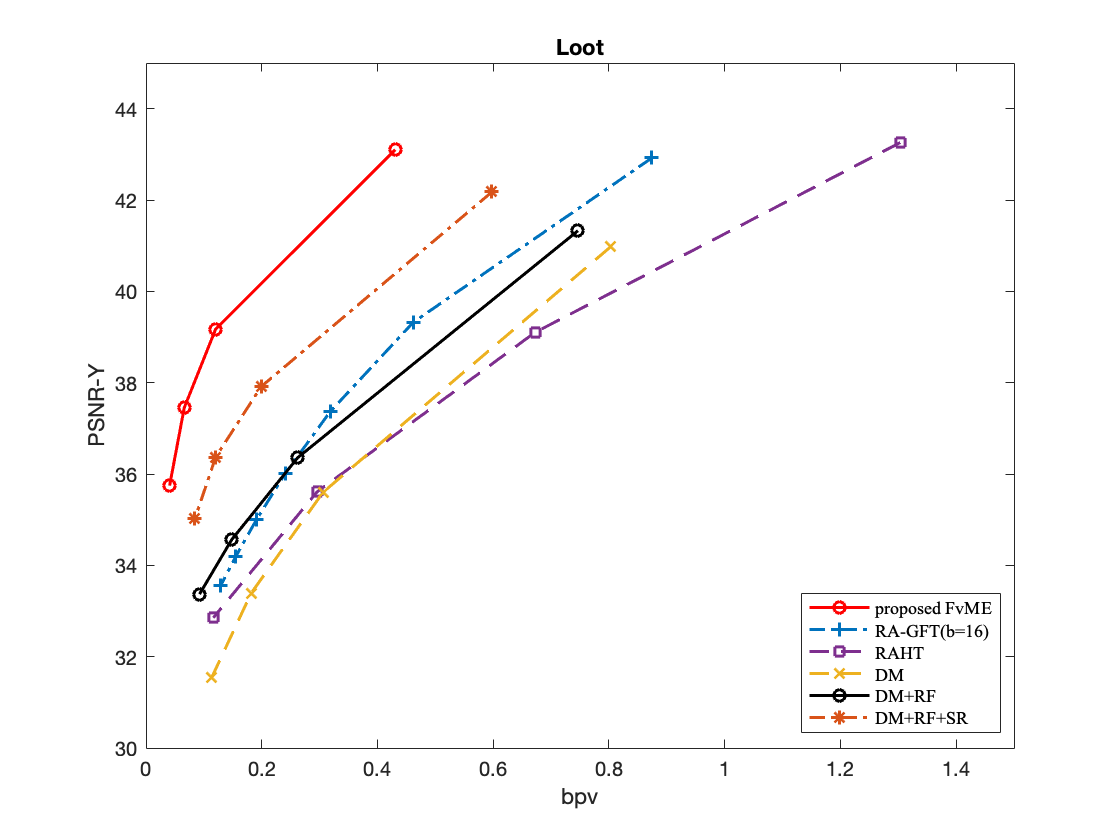}
\vspace{-6mm}
\caption{loot}
\end{subfigure}
\end{center}
\vspace{-5mm}
\caption{Rate distortion curves of  8iVFBv2 sequences.}
\label{fig:RDcurves}
\vspace{-2mm}
\end{figure}

Rate distortion (RD) curves are shown in Fig.~\ref{fig:RDcurves}. 
We first note that  using only the original  IvME from the database \cite{IMVdataset}, results in  sub-optimal performance compared to RAHT and RA-GFT. This is in part due to the criteria used in \cite{IMVdataset} to choose the optimal MV based on geometry and color information.  After local refinement with integer precision, the performance of IvME (DM+RF) improves significantly with respect to IvME (DM)  but it is still far from being competitive with other techniques.  Further improvements have been shown to be achievable by  using  per block intra/inter mode decision  \cite{Souto2020}.

\begin{table}[H]
\centering
\resizebox{0.8\linewidth}{!}{
\begin{tabular}{ |p{3.5cm} |p{1.5cm}|p{1.5cm} |p{1.5cm}|p{1.5cm} |}
\hline
 \textbf{\scriptsize anchors $\backslash$ sequences} & \textit{\scriptsize  longdress} & \textit{\scriptsize  soldier} & \textit{\scriptsize  redandblack} & \textit{\scriptsize  loot} \\
 \hline
 { \scriptsize {IvME(DM+RF)} } &
 { \scriptsize $-43.93\%$} & { \scriptsize $-63.96\%$} & { \scriptsize $-57.98\%$} & { \scriptsize $-64.43\%$}\\
  \hline
  { \scriptsize {RAHT} } &
 { \scriptsize $-39.94\%$} & { \scriptsize $-81.91\%$} & { \scriptsize $-51.10\%$} & { \scriptsize $-73.69\%$}\\
  \hline
  { \scriptsize {RA-GFT(b=16)} } &
 { \scriptsize $-16.19\%$} & { \scriptsize $-72.46\%$} & { \scriptsize $-24.37\%$} & { \scriptsize $-61.44\%$}\\
  \hline
\end{tabular}}
 \caption{The BD-rate performances of the proposed scheme over baselines} 
 \label{tab:BDBR}
\end{table}


After the reference blocks are super-resolved, the performance of the proposed IvME (DM+RF+SR) is further improved with respect to DM+RF, even without increasing MV resolution. The DM+RF+SR scheme can be better than the intra schemes in some cases with the advantage of complexity lower than that of the proposed FvME. Finally, after we increase MV resolution to 1/2-voxel, further coding gains are obtained, outperforming intra coding baselines,  RA-GFT and RAHT, with average gains of   $2.8dB$ and $4.6dB$, respectively. The method is always better than DM+RF+SR but at the cost of higher complexity due to additional motion search.
The results show that both interpolated fractional voxels and high resolution MVs lead to higher coding gain and outperform both inter coding with IvME and non predictive transform based schemes.

Table~\ref{tab:BDBR} summarizes the performance of the proposed method over IvME (DM+RF), RA-GFT, and RAGT in terms of BD-rate.  The proposed FvME can achieve $57\%$ average bitrate reduction over IvME (DM+RF). Compared with the prior arts, the proposed scheme can achieve $61\%$ and $43\%$ bitrate reduction on average over RAHT and RA-GFT respectively.

Compared with IvME (DM+RF), the proposed FvME scheme increases the number of voxels at most eightfold (because of super resolution), and   requires evaluating additional 27 fractional displacements for ME. Therefore, the complexity of the proposed FvME is larger than the complexity of IvME (DM+RF) by a constant factor (independent of the point cloud size). 
\section{Conclusions}
\label{sec:conclusion}
This paper describes a fractional-voxel motion estimation scheme tailored for attribute compression in 3D dynamic point clouds. Our scheme defines and identifies fractional voxels to be interpolated and provides a motion compensation prediction method by super-resolution and temporal correspondence. Extensive experiments show superior performance over the prior art. Our work reveals the benefits given by high resolution references and the further improvements given by fractional-voxel motion vectors for dynamic point clouds color compression. 

\section{References}
\bibliographystyle{IEEEbib.bst}
\bibliography{refs}
\end{document}